\documentclass{emulateapj}
\usepackage{ulem}
\usepackage{longtable}
\usepackage{graphicx}
\shorttitle{Fall-back accretion in GRB 121027A}

\begin{document}
\title{Giant X-ray Bump in GRB 121027A: Evidence for Fall-back Disk Accretion}
\author{Xue-Feng Wu$^{1,2,3}$, Shu-Jin Hou$^{4}$, and Wei-Hua Lei$^{5}$}
\altaffiltext{1}{Purple Mountain Observatory, Chinese Academy of Sciences, Nanjing 210008, China; xfwu@pmo.ac.cn}
\altaffiltext{2}{Chinese Center for Antarctic Astronomy, Chinese Academy of Sciences, Nanjing 210008, China}
\altaffiltext{3}{Joint Center for Particle Nuclear Physics and Cosmology of Purple Mountain Observatory and Nanjing University, Nanjing 210008, China}
\altaffiltext{4}{Department of Astronomy and Institute of Theoretical Physics and Astrophysics, Xiamen University, Xiamen, Fujian 361005, China}
\altaffiltext{5}{School of Physics, Huazhong University of Science and Technology, Wuhan, 430074, China; leiwh@hust.edu.cn}

\begin{abstract}
A particularly interesting discovery in observations of GRB 121027A is that of a giant X-ray bump detected by the {\it Swift}/XRT. The X-ray afterglow re-brightens sharply at $\sim 10^3$ s after the trigger by more than two orders of magnitude in less than 200 s. This X-ray bump lasts for more than $10^4$ s. It is quite different from typical X-ray flares. In this letter we propose a fall-back accretion model to interpret this X-ray bump within the context of the collapse of a massive star for a long duration gamma-ray burst. The required fall-back radius $\sim 3.5\times10^{10}$ cm and mass $\sim 0.9 - 2.6~M_\odot$ imply that a significant part of the helium envelope should  survive through the mass loss during the last stage of the massive progenitor of GRB 121027A.

\end{abstract}

\section{Introduction}

The most popular models of long-duration gamma-ray bursts (GRBs) invoke a collapse of a massive star (Woosley 1993; Paczynski 1998; MacFadyen \& Woosley 1999). The accretion of the stellar core by the central black hole (BH) fuels the prompt emission. However, the central engine activity does not cease after the prompt phase, which is especially supported by observations of late X-ray flares by the X-Ray Telescope (XRT) onboard the {\it Swift} satellite. X-ray flares share a lot of similarities with GRB prompt emission and  are therefore interpreted by the same mechanism as prompt emission, i.e., internal shocks (Burrows et al. 2005; Zhang et al. 2006; Fan \& Wei 2005). The fall-back and accretion of the stellar envelope may have some observational consequences (Kumar et al. 2008a, 2008b; Dai \& Liu 2012). In order to have a successive GRB jet penetrating the star, the progenitor is usually assumed to be a Wolf-Rayet star that has an evolved compact Helium envelope.

For the GRB BH central engine models, there are two main energy reservoirs to provide the jet power: the accretion energy in the disk which is carried by neutrinos and anti-neutrinos, that annihilate and power a bipolar outflow; and the spin energy of the BH which can be tapped by a magnetic field connecting the outer world through the Blandford-Znajek (1977, hereafter BZ) mechanism. Both models have been extensively investigated by many authors (e.g., Popham et al. 1999; Lee et al. 2000; Li 2000; Narayan et al. 2001; Di Matteo et al. 2002; Kohri \& Mineshige 2002; Wang et al. 2002; McKinney 2005; Gu et al. 2006; Chen \& Beloborodov 2007; Janiuk et al. 2007; Lei et al. 2009; Liu et al. 2007).

GRB 121027A may provide us a good chance to study the properties of GRB progenitor as well as the central engine models. As presented in the next section, the rising behavior of the giant X-ray bump in GRB 121027A is quite different with typical X-ray flares, so it should have different physical origin. In Section 2, we describe the prompt trigger and late XRT observations of GRB 121027A. In Section 3, we propose the fall-back accretion model and apply this model to the giant X-ray bump observed in GRB 121027A. In Section 4, We briefly summarize our results and discuss the implication.


\section{GRB 121027A Observations}

GRB 121027A was discovered at $T_0 = 07:32:29$ UT on 2012 October 27 by the Burst Alert Telescope (BAT) on board {\it Swift} (Evans et al. 2012a) and was later accurately located by XRT at a position of $\alpha$ = 04h 14m 23.37s, $\delta$ = -58$^{\circ}$49$'$46.6$''$ (J2000), with an uncertainty of $2.^{''}$0 (Beardmore et al 2012). The redshift of this burst is provisionally implied to be $z\sim 1.77$ by Gemini South spectroscopic observation, assuming the significant absorption feature at $7770{\rm \AA}$ in the optical spectrum as the MgII doublet $2796/2803{\rm \AA}$ (Tanvir et al. 2012). It is later confirmed and accurately measured to be $z=1.773$ by identifying several metal absorption lines with the X-shooter spectragraph (Kruehler et al. 2012). The results of data analysis of the prompt BAT observation are as follows (Barthelmy et al. 2012). The mask-weighted light curve shows a pulse with fast-rise and exponential-decay (FRED) profile plus two small peaks on the tail. The duration of GRB 121027A is $T_{90} = 62.6 \pm 4.8$ s in the $15 - 350$ keV band. The time-averaged spectrum is best fit by a simple power law with the photon index $\Gamma = 1.82 \pm 0.09$. The fluence in the $15 - 150$ keV band is $f_{\gamma}=2.0\pm0.1 \times 10^{-6}$ erg cm$^{-2}$ s$^{-1}$, yielding an isotropic gamma-ray energy release  $E_{\gamma,\rm iso}=4\pi D_{L}^2 f_{\gamma}/(1+z) = 1.58 \pm 0.08 \times10^{52}$ erg. Here we adopt the concordance cosmology with $\Omega_m=0.27$, $\Omega_{\Lambda}=0.73$ and $h_0=0.71$.

\begin{figure}[htc]
\center
\includegraphics[width=9cm,angle=0]{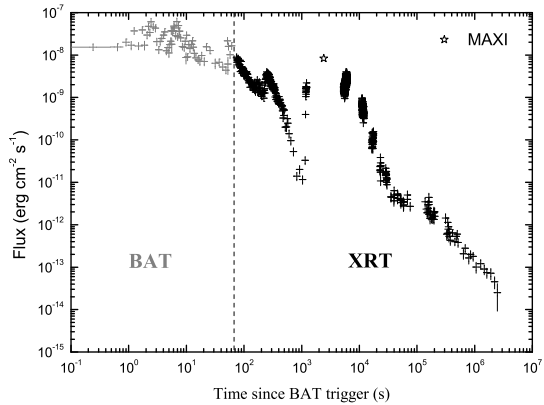}
\caption{BAT (gray) and XRT (black) light curves of GRB 121027A. BAT flux is calculated at 10 keV. XRT flux is absorption-corrected in the $0.3 - 10$ keV. The prompt and afterglow emission is separated by the dashed vertical line. Also plot is the Japanese MAXI/GSC observation (open pentagram), whose flux has been extrapolated from the $4 - 10$ keV band ($3.6\times 10^{-9}$ erg cm$^{-2}$ s$^{-1}$) to the $0.3 - 10$ keV band ($8.3\times10^{-9}$ erg cm$^{-2}$ s$^{-1}$).}
\label{fig1}
\end{figure}

The XRT began observing the burst at $T_0 + 67.4$ s. The XRT light curve shows several components (Evans et al. 2012b). Fig.\ref{fig1} shows the XRT light curve, which is the temporal evolution of the unabsorbed flux in the $0.3 - 10$ keV band (Evans et al. 2007, 2009). Initially it decays as a power law with the temporal index $\alpha_{1} \sim 1.8$ until an X-ray flare emerges at $T_0+220$ s and lasts for $\sim 300$ s. The decay slope of this X-ray flare is very steep, $\alpha_{2} \sim 6.6$.

The most interesting discovery in GRB 121027A is the giant X-ray bump in the subsequent observations. The flux of the giant X-ray bump increases sharply at $\sim T_0 + 10^3$ s by more than two orders of magnitude in less than 200 s. The flux is $1.2\times 10^{-11}$ erg cm$^{-2}$ s$^{-1}$ at $T_0+1033$ s (Windowed Timing mode), and suddenly increases to $1.7\times 10^{-9}$ erg cm$^{-2}$ s$^{-1}$ at $T_0+1198$ s (Photon Counting mode). Such ``step-like'' re-brightening of the X-ray bump in GRB 121027A is quite different with those of typical X-ray flares detected by {\it Swift} in the past eight years. For X-ray flares, the rise and decay time scales are compatible, which are close to the peak time of the flare (e.g., Liang et al. 2006; Chincarini et al. 2010). For the giant X-ray bump of GRB 121027A, the decay time scale, as shown in Figure 1, is $\sim 10^4$ s, much longer than the rising timescale. There is no XRT observation between $T_0+1.2$ ks and $T_0+5.3$ ks; however, the {\it MAXI}/Gas Slit Camera(GSC) detects the X-ray counterpart of GRB 121027A with the flux $\sim 150$ mCrab in the $4 - 10$ keV band at $T_0+2424$ s (Serino et al. 2012). Although the decay of the bump is not smooth and may have flaring features, its envelope can be roughly divided into two stages. Initially the decay slope is $\sim 1.6$ until $T_0+ 1.2\times10^{4}$ s. After that, the decay slope is $\sim 3.8$. The spectrum gets softening during the decay phase. The time-averaged photon index $\Gamma$ is $1.66\pm0.20$ between $T_0+5.3$ ks and $T_0+11.9$ ks, while it changes to a much softened value of $2.47\pm0.13$ between $T_0+16.3$ ks and $T_0+30$ ks. The late XRT emission is dominated by another component after $T_0+50$ ks, which can be attributed to the forward shock emission. More detailed data analysis of spectra and lightcurves of BAT and XRT emission of GRB 121027A can be found in Peng et al. (2013) and Levan et al. (2013).

\section{Fall-back Accretion Model}

\begin{figure}[htc]
\center
\includegraphics[width=7cm,angle=0]{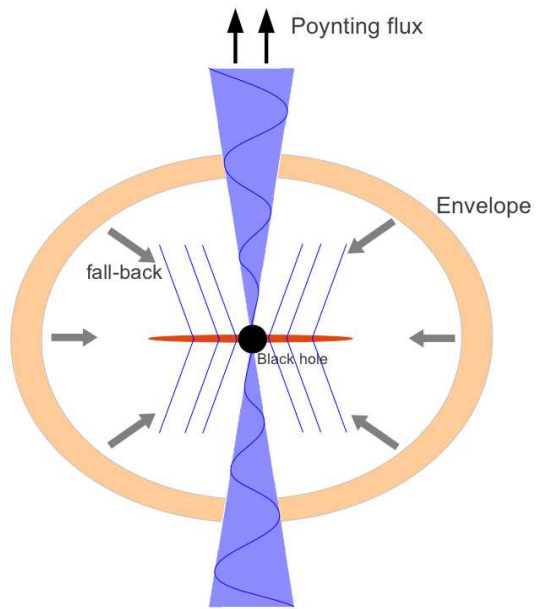}
\caption{Illustration of our model. The fall-back and accretion of the stellar envelope produce the X-ray bump seen in GRB 121027A. The jet is powered by the Blandford-Znajek mechanism, which extracts the rotational energy of Kerr BH through large-scale magnetic field (oblique and spiral lines).}
\label{fig2}
\end{figure}

We suggest that the X-ray bump seen in GRB 121027A is the result of the fall-back accretion, as shown in Fig.\ref{fig2}. The fall-back accretion is expected to start at the fall-back time $t_0 = t_{\rm fb}$ (in the following, time is defined in the cosmologically local frame). The fall-back time is the time it takes a parcel of gas of the progenitor star at radius $r_{\rm fb}$ to fall to the center, and it is approximately equal to the free-fall time, $t_{\rm fb} \sim (\pi^2 r_{\rm fb}^3 /8 G M_\bullet)^{1/2}$, where $M_\bullet$ is the BH mass.

Following MacFadyen et al. (2001), Zhang et al. (2008),and Dai \& Liu (2012), the fall-back accretion rate initially increases with time as $\dot{M}_{\rm early} \propto t^{1/2}$ until it reaches a peak value at $t_{\rm p}$. The late-time fall-back accretion behavior follows $\dot{M_{\rm late}}\propto t^{-5/3}$, as suggested by Chevalier (1989). Therefore, we assume that the fall-back accretion rate evolves as a smooth-broken-power-law function of time,\footnote{The actual peak time and peak accretion rate in Eq. (1) are slightly different from $t_{\rm p}$ and $\dot{M}_{\rm p}$.}
\begin{equation}
\dot{M} = \dot{M}_{\rm p} \left[ \frac{1}{2}\left(\frac{t-t_0}{t_{\rm p}-t_0} \right)^{-\alpha_r s} +  \frac{1}{2}\left(\frac{t-t_0}{t_{\rm p}-t_0} \right)^{-\alpha_d s} \right]^{-1/s}
\label{eq:dotM}
\end{equation}
where $\alpha_r=1/2$, $\alpha_d=-5/3$ and $s$ describes the sharpness of the peak. The dimensionless accretion rate is defined as $\dot{m} = \dot{M}/(M_\odot /\rm s)$.

As suggested by Lei et al. (2013; see also Lei et al. in preparation), the jet may be dominated by the BZ power especially at late time (Fan et al. 2005; Zhang \& Yan 2011). For this reason, we connect the observed X-ray luminosity to the BZ power through
\begin{equation}
\eta \dot{E}_{\rm B}=f_{\rm b} L_{\rm X,iso}
\label{eq1}
\end{equation}
where $\eta$ is the efficiency of converting BZ power to X-ray radiation and $f_{\rm b}$ is the beaming factor of the jet.

The BZ jet power from a BH with mass $M_{\bullet}$ (or dimensionless mass $m_{\bullet}=M_{\bullet}/M_\odot$) and angular momentum $J_\bullet$ is (Lee et al. 2000; Li 2000; Wang et al. 2002; McKinney 2005; Lei \& Zhang 2011; Lei et al. 2013)
\begin{equation}
\dot{E}_{\rm B}=1.7 \times 10^{50} a_{\bullet}^2 m_{\bullet}^2
B_{\bullet,15}^2 F(a_{\bullet}) \ {\rm erg \ s^{-1}},
\label{eq_Lmag}
\end{equation}
where $B_{\bullet,15}=B_{\bullet}/10^{15} {\rm G}$ and
\begin{equation}
F(a_{\bullet})=[(1+q^2)/q^2][(q+1/q) \arctan q-1].
\label{eq_F}
\end{equation}
Here $a_\bullet = J_\bullet c/(GM_\bullet^2)$ is the BH spin parameter, and $q= a_{\bullet} /(1+\sqrt{1-a^2_{\bullet}})$. For
$0\leq a_{\bullet} \leq 1$, $2/3\leq F(a_{\bullet}) \leq \pi-2$. It is obvious that the BZ power depends on $M_{\bullet}$, $B_{\bullet}$,
and $a_{\bullet}$. A strong magnetic field of $\sim 10^{15} \rm G$ is required to produce the high luminosity of a GRB.

As the magnetic field on the BH is supported by the surrounding disk, there are some relations between $B_{\bullet}$ and $\dot{M}$.  In a hyper-accreting flow in a GRB, it is possible that a magnetic flux is accumulated near the BH horizon. Considering the balance between the magnetic pressure on the horizon and the ram pressure of the innermost part of the accretion flow (e.g. Moderski et al. 1997), one can estimate the magnetic field strength threading the BH horizon
\begin{equation}
\frac{B_{\bullet}^2}{8\pi} = P_{\rm ram} \sim \rho c^2 \sim \frac{\dot{M} c}{4\pi r_{\bullet}^2}
\label{Bmdot}
\end{equation}
where $r_{\bullet}=(1+\sqrt{1-a_\bullet^2})r_{\rm g}$ is the radius of the BH horizon, and $r_{\rm g} = G M_\bullet /c^2$. It can be rewritten as
\begin{equation}
B_{\bullet} \simeq 7.4 \times 10^{16} \dot{m}^{1/2} m_{\bullet}^{-1} (1+\sqrt{1-a_\bullet^2})^{-1} \rm{G}.
\label{eq:B}
\end{equation}
Inserting it to Equation (\ref{eq_Lmag}), we obtain the magnetic power as a function of mass accretion rate and BH spin, i.e.
\begin{equation}
\dot{E}_{\rm B}=9.3 \times 10^{53} a_\bullet^2 \dot{m}  X(a_\bullet) \ {\rm erg \ s^{-1}} ,
\label{eq:EB}
\end{equation}
and
\begin{equation}
X(a_\bullet)=F(a_\bullet)/(1+\sqrt{1-a_\bullet^2})^2.
\end{equation}
It is found that $X(0)=1/6$, and $X(1)=\pi -2$. In general, a faster BH is more favorable for GRB production, as w revealed  by recent GRMHD numerical simulations (Nagataki 2009).

The BH evolves with time during a GRB, since it would be spun up by accretion and spun-down by the BZ mechanism. The evolution equations of a Kerr BH in the BZ model can be written as
\begin{equation}
\frac{dM_\bullet c^2}{dt} = \dot{M} c^2 E_{\rm ms} - \dot{E}_{\rm B},
\label{eq:dMbz}
\end{equation}

\begin{equation}
\frac{dJ_\bullet}{dt} = \dot{M} L_{\rm ms} - T_{\rm B}.
\label{eq:dJbz}
\end{equation}
Since $a_\bullet = J_\bullet c/(GM_\bullet^2)$, by incorporating Eqs. (\ref{eq:dMbz}) and (\ref{eq:dJbz}), the evolution of the BH spin can be expressed by
\begin{eqnarray}
\frac{da_\bullet}{dt} = && (\dot{M} L_{\rm ms} - T_{\rm B})c/(GM_\bullet ^2) - \nonumber \\
&& 2 a_\bullet (\dot{M} c^2 E_{\rm ms} - \dot{E}_{\rm B}) /(M_\bullet c^2),
\label{eq:da}
\end{eqnarray}
where $E_{\rm ms}$ and $L_{\rm ms}$ are the specific energy and angular momentum corresponding to the inner most radius $r_{\rm ms}$ of the disk, respectively, which are defined as (Novikov \& Thorne 1973)
\begin{equation}
E_{\rm ms} = \frac{4\sqrt{ R_{\rm ms} }-3a_{\bullet}}{\sqrt{3} R_{\rm ms}},
\end{equation}
\begin{equation}
L_{\rm ms} = \frac{G M_\bullet}{c} \frac{2 (3 \sqrt{R_{\rm ms}} -2 a_\bullet) }{\sqrt{3} \sqrt{R_{\rm ms}} },
\end{equation}
where $R_{\rm ms} = r_{\rm ms}/r_{\rm g}$ is the radius of the marginally stable orbit in terms of $r_{\rm g}$. The expression for $R_{\rm ms}$ is (Bardeen et al. 1972),
\begin{eqnarray}
R_{\rm ms} =  3+Z_2 -\left[(3-Z_1)(3+Z_1+2Z_2)\right]^{1/2},
\end{eqnarray}
for $0\leq a_{\bullet} \leq 1$, where $Z_1 \equiv 1+(1-a_{\bullet}^2)^{1/3} [(1+a_{\bullet})^{1/3}+(1-a_{\bullet})^{1/3}]$, $Z_2\equiv (3a_{\bullet}^2+Z_1^2)^{1/2}$.

In Eq. (\ref{eq:dJbz}), $T_{\rm B}= {\dot{E}_{\rm B}}/{\Omega_{\rm F}}$ is the total magnetic torque applied on the BH, i.e.


\begin{eqnarray}
T_{\rm B} =3.36 \times 10^{45} a_\bullet^2 q^{-1} m_{\bullet}^3 B_{\bullet,15}^2  F(a_\bullet){\rm \ g \ cm^2 \ s^{-2}}. \nonumber \\
\end{eqnarray}

Here $\Omega_{\rm F}=0.5\Omega_\bullet$ is usually taken to maximize the BZ power, and
\begin{equation}
\Omega_\bullet = \frac{c^3}{G M_\bullet} \frac{a_\bullet}{2 (1+\sqrt{1-a_\bullet^2})}
\end{equation}
is the angular velocity of the BH horizon.

One can calculate the evolution of the BH spin by inserting the above expressions into Eq.(11), and then study the time-dependent X-ray luminosity by substituting the evolutions of the BH sin and fallback accretion rate into Eq.(\ref{eq:EB}) of the BZ power.

For GRB 121027A, the X-ray bump appears at $\sim T_0 + 1000$ s after the GRB trigger, which, divided by $1+z$, corresponds to $t_{\rm fb}\sim t_0\sim 360$ s. This suggests the minimum radius around which matter starts to fall back is
\begin{equation}
r_{\rm fb} \simeq 3.5 \times 10^{10} (M_\bullet /3 M_\sun)^{1/3} ({t_{\rm fb}/360\;\rm s})^{2/3} \rm cm.
\end{equation}

The peak flux of the bump in the $1-10$ keV band is $F_{\rm X,peak} > 8.3 \times 10^{-9}\; {\rm erg ~
cm^{-2} ~ s^{-1}}$, corresponding to a peak luminosity $L_{\rm X,iso}^{\rm peak} > 1.8 \times 10^{50}~{\rm erg~s^{-1}}$. From Eq. (\ref{eq:EB}), we can estimate the peak accretion rate
\begin{equation}
\dot{M}_{\rm p} \simeq 1.1 \times 10^{-4} L_{\rm X,iso,50} a_\bullet^{-2} X^{-1}(a_\bullet)  \eta_{-2}^{-1} f_{\rm b,-2} M_{\sun}~{\rm s}^{-1},
\end{equation}
where $\eta_{-2} =\eta/10^{-2}$ and $f_{\rm b,-2}=f_{\rm b}/10^{-2}$.

\begin{figure}[htc]
\center
\includegraphics[width=10cm,angle=0]{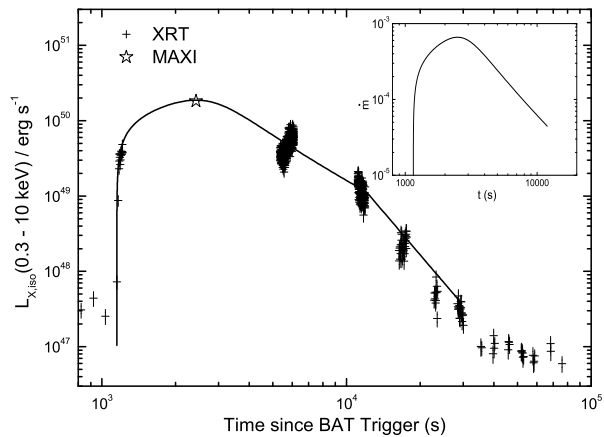}
\caption{Fitting (solid line) to the X-ray bump of GRB 121027A ($a_\bullet=0.9$). The inset panel shows the evolution of the dimensionless mass accretion rate.}
\label{fig3}
\end{figure}

The total rising time of the bump is about $1800$ s (Fig. 3), we thus have $t_{\rm p}-t_0 \sim 1800/(1+z)$ s $\sim 650$ s. By Eq.(\ref{eq:dotM}), the total fallback/accreted mass should be
\begin{eqnarray}
M_{\rm fb}&& \simeq \int_{t_0}^{t_p} \dot{M}dt \sim 2\dot{M}_{\rm p} (t_{\rm p} -t_0)/3  \nonumber \\
 && \simeq  4.6 \times 10^{-2} L_{\rm X,iso,50} a_\bullet^{-2} X^{-1}(a_\bullet)  \eta_{-2}^{-1} f_{\rm b,-2} M_{\sun}. \nonumber \\
\end{eqnarray}

From Eq.(\ref{eq:B}), the maximum magnetic field strength around BH is
\begin{equation}
B_{\bullet, \rm p} \simeq 7.8 \times 10^{14} L_{\rm X,iso,50}^{1/2} m_\bullet^{-1} q a_\bullet^{-2} X^{-1/2}(a_\bullet) \eta_{-2}^{-1/2} f_{\rm b,-2}^{1/2} \rm G.
\end{equation}
The exact values of the above parameters depend strongly on the BH spin $a_\bullet$ at time $t_{\rm p}$.

To constrain the above parameter values in GRB 121027A, we carried out numerical calculation of Eqs.(\ref{eq:dotM}) - (\ref{eq:da}). We obtain the time evolution of the BZ power, and compare it with the observations of the X-ray bump in GRB 121027A. In our calculation, we assume $\eta=10^{-2}$, $f_{\rm b}=10^{-2}$ (the jet half-opening angle is constrained to be $\theta_{\rm j}>0.2$ radian by late XRT observation; Peng et al. 2013). The BH is initially set up with a mass $m_\bullet=3$ and a spin $a_\bullet=0.9$. The calculation starts at $t_0=1150/(1+z)$ s. Fig.\ref{fig3} shows our model fit to the X-ray bump in GRB 121027A. The parameters for the fitting are $\dot{m}_{\rm p} = 6.1 \times 10^{-4}$, $s=1.9$,and $t_{\rm p}=2950/(1+z)$ s. The total fall-back mass is $M_{\rm fb}= 0.9 M_\sun$. The total fallback mass is one order of magnitude higher than that estimated with Eq. (19), which underestimates the actual duration of smoothed peak accretion (see Fig. 3). The initial spin of the black hole for the X-ray bump phase has large uncertainty. We also considered low spin cases. For $a_\bullet=0$, the model parameters are $\dot{m}_{\rm p} = 7.0 \times 10^{-3}$, $s=0.35$, $t_{\rm p}=1350/(1+z)$ s, and $M_{\rm fb}=2.6 M_\sun$. For $a_\bullet=0.5$, the model parameters are $\dot{m}_{\rm p} = 1.7 \times 10^{-3}$, $s=0.70$, $t_{\rm p}=2050/(1+z)$ s, and $M_{\rm fb}=1.8 M_\sun$. Note that for the decay phase of the bump, we focus on the temporal evolution of the envelope of XRT emission. There may be fragmentation during the fall-back phase (King et al. 2005), which can account for the variations of XRT flux during the decay phase.

There is a break in the lightcurve at time $t_{\rm b}\sim 1.2\times 10^4 \rm s$, after which the photon index $\Gamma$ changes from $1.66\pm0.20$ to $2.47\pm0.13$. If the central engine ceases at this time, one can expect a transition in the lightcurve from the fallback phase to the tail emission phase. In this scenario, the temporal decay index after $t_{\rm b}$ is $\alpha=1+\Gamma =3.47\pm0.13$ due to the curvature effect, which is consistent with the observed one ($3.8$). It requires that the mass fallback should stop at $t_{\rm b}$. Therefore, remnant emission comes from high latitude and the observed XRT flux fades as $t^{-3.8}$ for $t>t_{\rm b}$.


\section{Conclusions and Discussion}

The X-ray afterglow light curve of GRB 121027A is unusual. The ``step-like'' re-brightening at about 1000 s since the burst with a duration longer than $10^4$ s,which we refer to as the giant X-ray bump in this Letter, is quite different with typical X-ray flares observed by {\it Swift}. We propose a fall-back accretion model to interpret this X-ray bump within the context of the collapse of a massive star for long duration GRBs. The fallback radius of $r_{\rm fb}\sim 3.5\times10^{10}$ cm and mass $M_{\rm fb}\sim 0.9 - 2.6~M_\odot$ for this burst require the helium envelope of the progenitor should be partly survived before the ending of the massive star. One may ask why this burst shows the fall-back signature, while most other long GRBs do not. One should always have fall-back. The reason may be that in the collapsar models, the bounding shock responsible for the associated supernova transfers kinetic energy to the envelope materials. The more energetic the supernova shock, the less envelope material falls back into the center. The potential energy of the fall-back material at $r_{\rm fb}$ is negative, with an  absolute value is $ GM_\bullet M_{\rm fb}/r_{\rm fb} \sim 2\times 10^{49}(M_{\rm fb}/M_\odot)$ erg for GRB 121027A, assuming $M_\bullet\sim 3 M_\odot$. If the kinetic energy delivered from the supernova shock is less than the potential energy, then these material will fall back. In our scenario, GRB 121027A might be accompanied with a low energy supernova, or even a failed supernova.

In this work, we do not include the magnetic coupling effect (see Eqs. \ref{eq:dMbz} and \ref{eq:dJbz}) between the BH and the disk through closed magnetic field lines (Li \& Paczynski 2000; Wang et al. 2002; Lei et al. 2009; Janiuk \& Yuan 2010). Similar to the BZ mechanism, the magnetic coupling effect also extracts rotational energy from the spinning BH. Only if the BH spin is initially small would the magnetic coupling  act as an additional spin-up process. A similar discussion on this aspect was made by Dai \& Liu (2012) within the context of the magnetar central engine model. In more general cases, the magnetic coupling effect would not significantly affect the BH spin evolution (Lei et al. 2009).

\acknowledgements We thank the referee for helpful suggestion in improving this Letter. We also thank Zigao Dai, Yizhong Fan, Enwei Liang, Jufu Lu, and Bing Zhang for helpful discussion. This work made use of data supplied by the UK Swift Science Data Centre at the University of Leicester. This work was supported by National Basic Research Program (``973'' Program) of China under Grant Nos. 2009CB824800 and 2013CB834900,the National Natural Science Foundation of China (grants 11233006, 11003004, 11173011 and U1231101). X-F W. acknowledges support by the One-Hundred Talents Program and the Youth Innovation Promotion Association of Chinese Academy of Sciences.


\end{document}